# Comment on "Comment on 'Spin-Momentum-Locked Edge Mode for Topological Vortex Lasing, Phys. Rev. Lett. vol. 125, 013903 (2020)'"

Zhen-Qian Yang,[1,2] Zeng-Kai Shao,[1,2] Hua-Zhou Chen,[1,2] Xin-Rui Mao,[1,2] and Ren-Min Ma,[1,2*]

[1]*State Key Lab for Mesoscopic Physics and School of Physics, Peking University, Beijing, China*
[2]*Frontiers Science Center for Nano-optoelectronics & Collaborative Innovation Center of Quantum Matter, Beijing, China*
*Email: renminma@pku.edu.cn

In our Letter [1], we reported topological vortex lasers based on spin-momentum-locked edge modes. We observed that the near field spin and orbital angular momentum has a one-to-one far-field radiation correspondence of circular polarization and orbital angular momentum respectively. Sun et al. in their Comment [2], however, argued that we did not perform numerical simulations on the near field information of our experimentally studied topological edge modes, and our mode assignment was mistaken and spoiled the one-to-one correspondence. However, we will show that their arguments are wrong. Furthermore, we will show that the Eqs. (1) and (2) and the phase windings in their Comment [2] are wrong.

We have performed numerical simulations, based on which we obtained Fig. 3b in our Letter and Fig S9 in its Supplemental Material. On page 3 of the main text, we specified, "Fig. 3(b) shows the full wave simulated pattern of $|-2,+\rangle$ mode, which matches well with the experimental one". In the caption of Fig. S9, we specified, "Simulated distribution of the dominant in-plane Ex field of an edge state". In Part 5 of numerical simulation of Supplemental Material, we specified our numerical simulation method. We need to highlight again that, we have performed numerical simulations on the near field information of our experimentally studied topological edge modes, and the results match very well with our experimental results.

There is no mistaken assignment of the modes in our Letter. The one-to-one correspondence has been verified in both of simulation and experiment. In their comment, the authors give the phase windings of $6\pi$ and $-2\pi$ to $|2,+\rangle$ and $|-2,+\rangle$ modes around the cavity respectively. However, this phase windings are wrong by messing up spin and orbital angular momenta. A topological edge mode consists of both dipole and quadruple modes [3-5]. The phase factors before $|p_\pm\rangle$ and $|d_\pm\rangle$ are $e^{il\varphi}$ and $e^{i(l\mp 1)\varphi}$ respectively (for instance, using their Eqs. (1) and (2)), which directly gives the phase windings of dipole and quadruple modes around the cavity of $2l\pi$ and $2(l\mp 1)\pi$ respectively. So, for the $|2,+\rangle$ mode, the phase windings of dipole and quadruple modes should be $4\pi$ and $2\pi$ respectively; for the $|-2,+\rangle$ mode, the phase windings of dipole and quadruple modes should be $-4\pi$ and $-6\pi$ respectively.

Their assignment of $|2,+\rangle$ and $|-2,+\rangle$ with $6\pi$ and $-2\pi$ phase windings around the cavity is not physically meaningful in both near field and far field. More importantly, messing up spin and orbital angular momenta ruins the physics of spin-momentum-locked edge modes and the one-to-one correspondence from near field to far field. Taking $|2,+\rangle$ and $|-2,+\rangle$ modes as an example, in the near field, $l = 2,$ or $-2$ means the relative phase of the $|p_+\rangle$ mode changes $4\pi$ or $-4\pi$ around one circle of the cavity. The radiation of a topological edge mode is dominated by its dipole mode component radiation, because quadruple modes are dark modes [6]. In the far field of $|2,+\rangle$ and $|-2,+\rangle$ modes, their near field $|p_+\rangle$ convert to circular polarization, while $|p_+\rangle$ phase windings around the cavity convert to opposite topological charges of 2 and −2 with $4\pi$ and $-4\pi$ phase windings around the singular point respectively, where one-to-one correspondence is obvious.

The Eqs. (1) and (2) in their Comment are wrong. There is no $\frac{1}{2}\delta\omega$ frequency shift in the expressions of their cavity resonant frequency. Because of the finite cavity size, for any cavity mode, there will be an additional $\frac{1}{2}\delta\omega$ frequency shift from the Dirac point in the dispersion curve of edge modes of an open topological waveguide with infinite size. This $\frac{1}{2}\delta\omega$ frequency shift is not trivial when comparing the resonant frequency of a cavity mode to dispersion curve of edge modes in an open topological waveguide with infinite size. As illustrated in our Letter and in its Supplemental Material, Fig. 1(c) and Fig. S6 are intended to highlight the discrete nature of the edge modes due to the finite cavity size, and to highlight their observable frequency correspondences to the experimental results and the calculated dispersions of bulk and edge modes with infinite size. To avoid any possible misunderstanding of the Fig. 1c and Fig. S6, we like to add the following information in the caption of the Fig. 1c and Fig. S6, which reads, "The horizontal coordinate does not apply to the cavity modes, whose momenta are given by their discrete orbital angular momentum and spin".